\documentclass{article}

\usepackage[final]{mytemplate}
\usepackage{pdflscape}
\usepackage{rotating}

\usepackage[utf8]{inputenc} 
\usepackage[T1]{fontenc}    
\usepackage{hyperref}       
\usepackage{url}            
\usepackage{booktabs}       
\usepackage{amsfonts}       
\usepackage{nicefrac}       
\usepackage{microtype}      
\usepackage{hyperref}
\usepackage{bbm}
\usepackage{mathtools}      
\usepackage{import}
\usepackage{csvsimple} 
\usepackage{caption} 
\usepackage{wrapfig}
\usepackage{tabularx}
\usepackage{multirow}

\usepackage[space]{cite}
\usepackage{amsmath}
\usepackage{amssymb}
\usepackage{amsthm}
\usepackage{microtype}
\usepackage{graphicx}
\usepackage{subfigure}
\usepackage{dsfont}
\usepackage{booktabs} 
\usepackage{adjustbox}
\usepackage{diagbox} 
\usepackage{etoolbox}
\usepackage{scalerel,stackengine}
\usepackage{bbm}
\usepackage{algorithm}
\usepackage{algpseudocode}

\usepackage{listings}
\usepackage{xparse}
\usepackage{xcolor}
\usepackage{epstopdf}
\usepackage{epsfig}
\usepackage{enumitem}

\stackMath
\newcommand\autowidehat[1]{%
\savestack{\tmpbox}{\stretchto{%
  \scaleto{%
    \scalerel*[\widthof{\ensuremath{#1}}]{\kern0pt\bigwedge\kern0pt}%
    {\rule[-\textheight/2]{1ex}{\textheight}}
  }{\textheight}%
}{0.5ex}}%
\stackon[1pt]{#1}{\tmpbox}%
}

\usepackage{mathtools}
\newcommand{\defeq}{\vcentcolon=}

\renewcommand{\S}{\mathbf{S}}
\newcommand{\m}{\mathbf{m}}

\newcommand{\Z}{\mathbf{Z}}

\newenvironment{talign*}
 {\csname align*\endcsname}
 {\endalign}
\newenvironment{talign}
{\align}
{\endalign}

\setlist[itemize]{leftmargin=5mm,itemsep=0.5mm}
\setlist[enumerate]{leftmargin=*,itemsep=0.5mm}
\title{Aggregated Gaussian Processes with Multiresolution Earth Observation Covariates}
\author{
  Harrison Zhu, Adam Howes  \\
  Imperial College London\\
  \texttt{\{harrison.zhu15, adam.howes19\}@imperial.ac.uk} 
  \and \textbf{Owen van Eer}, \textbf{Maxime Rischard}\\
   Cervest\\
  \texttt{\{owen, maxime\}@cervest.earth}
  \and
   \textbf{Yingzhen Li} \\
  Imperial College London \\
  \texttt{yingzhen.li@imperial.ac.uk}
  \and
   \textbf{Dino Sejdinovic} \\
  University of Oxford \\
  \texttt{dino.sejdinovic@stats.ox.ac.uk}
  \and
  \textbf{Seth Flaxman} \\
  University of Oxford \\
  \texttt{seth.flaxman@cs.ox.ac.uk} \\
}

\begin{document}
\maketitle

\begin{abstract}
    For many survey-based spatial modelling problems, responses are observed as spatially aggregated over survey regions due to limited resources. Covariates, from weather models and satellite imageries, can be observed at many different spatial resolutions, making the pre-processing of covariates a key challenge for any spatial modelling task. We propose a Gaussian process regression model to flexibly handle multiresolution covariates by employing an additive kernel that can efficiently aggregate features across resolutions. Compared to existing approaches that rely on resolution matching, our approach better maintains distributional information across resolutions, leading to better performance and interpretability. Our model yields stronger predictive performance and interpretability on both simulated and crop yield datasets. 
    \end{abstract}
    
    \section{Introduction}
    For many problems in sustainable development and public health, the abundance of Earth observation data, such as satellite imagery, has helped infer the underlying trends and guide policy decisions. However, taking crop yield modelling as an example, crop yields are only observed yearly at the county level due to the scarcity of agricultural census data \citep{burke_using_2021}, whereas Earth observation covariates are abundantly available at different temporal (e.g.~weekly) and  pixel-level (e.g.~250m by 250m pixels in space for the MODIS satellite products \citep{didan_mod13q1_2015}) spatial resolutions. Similarly, in public health, disease outcome data \citep{bhatt_improved_2017,lucas_improving_2020, arambepola2020simulation} is also often only surveyed at the census (or aggregated) level due to privacy and limited resources. To accomplish modelling at the census level, a straightforward and widely-used approach is to spatially aggregate or average the covariates within each census level in order to create a standard supervised learning dataset \citep{you_deep_2017,bhatt_improved_2017,mateo-sanchis_synergistic_2019, fan2021a}. The drawback of this approach is that it results in a loss of within region-level variability and hinders pixel-level prediction \citep{law_variational_2018, lucas_improving_2020, arambepola2020simulation,stefanovic2021reconstructing} due to fine-scale covariate information being discarded. For many applications, high-resolution predictions can be very important, e.g. for policy making purposes.
    
    
    Crop yield and disease modelling problems often are \textbf{multiple instance regression} problems 
    \begin{talign*}
        y_i = g(\mathbf{X}^1_i, \ldots,\mathbf{X}^d_i) + \epsilon_i, \quad \epsilon_i\sim\text{a noise distribution},
    \end{talign*} 
    where the responses $y_i$ are available at a much lower spatial resolution than the covariates $\mathbf{X}^1_i\in\mathbb{R}^{N_{i1}\times D_1}, \ldots,\mathbf{X}_i^d\in\mathbb{R}^{N_{id}\times D_d}$ and where each of the covariates may have a different \textbf{spatial resolution}, represented by the number of \textbf{pixels} $N_{il}$ for $l=1,\ldots,d$ resolutions, and \textbf{number of dimensions} $D_l$. Aggregated Gaussian processes \citep{law_variational_2018, yousefi_multi-task_2019,hamelijnck_multi-resolution_2019,tanaka_spatially_2019, lucas_improving_2020, arambepola2020simulation} define the mapping $g$ with an aggregation, which is a linear operator, $\text{Agg}_i(f)=\int_{\mathcal{X}} f(x) \text{d}\Pi_i(x)$ to a Gaussian process prior $f$ with a distribution $\Pi_i$ over the covariates living in $\mathcal{X}$. With $\mathbf{X}_i=[\mathbf{X}^1_i, \ldots,\mathbf{X}^d_i]$, assuming that $N_i:=N_{i1}=\cdots=N_{id}$ (different resolutions can be \textit{matched} to a single resolution) and $\mathbf{X}_{ij}\equiv x_{ij}\stackrel{iid}{\sim}\Pi_i$, one option is to use Monte Carlo integration to get $\text{Agg}_i(f)\approx N_{i}^{-1}\sum_{j=1}^{N_{i}} f(x_{ij})$.
    
    Aggregated Gaussian processes have been successfully applied to a variety of relevant problems, such as malaria prevalence mapping \citep{law_variational_2018, lucas_improving_2020,arambepola2020simulation} and air pollution mapping \citep{hamelijnck_multi-resolution_2019}, both for probabilistic prediction and explaining real world phenomena to guide policy. The latter can be achieved by predicting for $f(x)$ at the pixel level, known as \textbf{disaggregation}. Existing works use sparse variational Gaussian process (SVGP; \citet{hensman_gaussian_2013,hensman_scalable_2015}) to reduce the computational complexity of inference from $\mathcal{O}(\sum_{i,j=1}^n N_{i} N_{j} + n^3)$ to $\mathcal{O}(M^3 + M^2\sum_{i=1}^{n_b} N_{i})$, where $n_b$ is the minibatch size and $M$ is the number of inducing points. 
    
    We propose a multiresolution aggregated Gaussian processes model that can flexibly handle the multiresolution nature of Earth observation data using additive kernels:
    \begin{itemize}
        
        \item Previous approaches focus on the single resolution case, thereby requiring data preprocessing/aggregation to match the resolutions. Just like original aggregated Gaussian processes circumvent the coarsening of covariates to the level of response, in order to preserve more granular information in the data, we minimise the need to preprocess multiple covariates into the same resolution - hence keeping more available covariate information intact across resolutions.
        
        \item Through a simulation study that mimics real world data and a USA soybeans dataset, we demonstrate that multiresolution modelling is indeed important in practice and that it results in the gains in both predictive performance and model interpretability. Through these numerical experiments, we thereby also provide a workflow to demonstrate how to deal with data preprocessing of different resolutions in practice. 
    \end{itemize}
    
    \section{Background}
    
    
    \paragraph{Gaussian Processes over Distributions:} 
    \begin{figure}[t]
        \centering
        \vspace{-3mm}
        \includegraphics[width = \columnwidth]{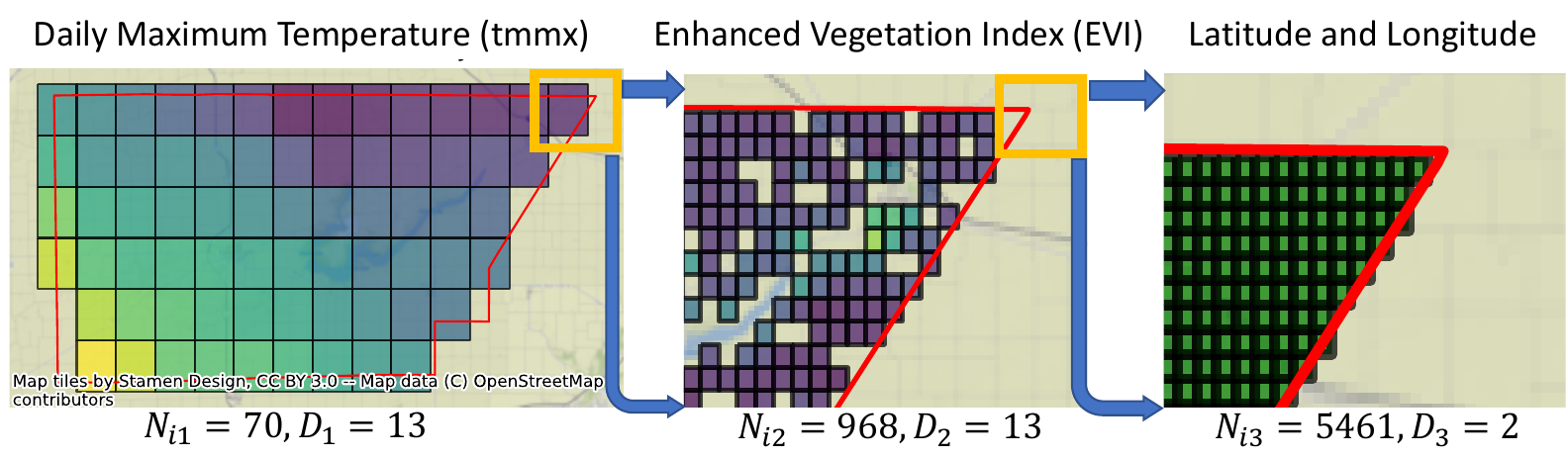}
        \vspace{-5mm}
        \caption{Visualisation of different resolutions within a county. The number of pixels $N_{il}$ vary between resolutions $l$.}
        \label{fig:covariates_layered}
    \end{figure}
    
    Taking the enhanced vegetation index (EVI) from the MODIS \citep{didan_mod13q1_2015} satellite system, observed over a county, as an example, the EVI is observed every 13 days at 250m spatial resolution, and as illustrated in Figure~\ref{fig:covariates_layered}, each county will contain a distinct number of pixels spatially. It is therefore natural to model county level responses based on the similarity between the sets of pixels in different counties. Although the covariates are available as sets, we would like to see them as finite samples from a distribution. Intuitively, if we could measure at infinitely high resolution, then under the infinite limit, the empirical distribution of the set would converge to a distribution. An alternative approach would be to directly model the sets, as proposed in Deep Sets \citep{zaheer2017deep}, but these have not shown to be competitive compared to kernel methods \citet{lemercier_distribution_2021}, which we will use.
    
    We assume that the data comes in the form of $\{(\Pi_i, y_i)\}_{i=1}^n$, where $y_i\in\mathbb{R}$ is a label, $\Pi_i$ is a distribution over covariates $x\in\mathcal{X}$, where $\mathcal{X}$ is a Banach space e.g. $\mathcal{X}\subseteq\mathbb{R}^d$ for $d\in\mathbb{N}_+$ \citep{law_variational_2018} or $d$-dimensional time series \citep{lemercier_distribution_2021}. As in \citet{lemercier_distribution_2021}, define $\mathcal{P}(\mathcal{X})$ to be the space of distributions on $\mathcal{X}$. We focus on probability measures $\mathcal{P}(\mathcal{X})$ but note that arguments with finite measures are almost identical.
    
    We denote $\mathcal{N}(a,b)$ as the distribution and $N(\cdot;a,b)$ the density function of a Gaussian distribution with mean $a$ and covariance $b$. In practice, $\Pi_i$ is not known explicitly and is estimated with the empirical measure $\hat{\Pi}_i=N_i^{-1}\sum_{j=1}^{N_i} \delta_{x_{ij}}$ for a \textbf{bag} of pixels $\{x_{ij}\}_{j=1}^{N_i}$, where $\delta_{x_{ij}}$ is a Dirac measure centred at $\{x_{ij}\}$ and $N_i\in\mathbb{N}_+$. Alternatively, one also could use survey weights \citep{law_variational_2018} or model the weights using stochastic processes \citep{hamelijnck_multi-resolution_2019}.


    
    Gaussian processes \citep{rasmussen_gaussian_2003} have proven to be a versatile family of probabilistic models for a variety of problems in spatial statistics \citep{gelfand_handbook_2010}. We can define a Gaussian process over distributions as $f\sim\mathcal{GP}(0, \rho)$ with a kernel $\rho:\mathcal{P}(\mathcal{X})\times \mathcal{P}(\mathcal{X})\rightarrow\mathbb{R}$ that is positive definite. Just as regular Gaussian processes, for any $\Pi_1,\ldots,\Pi_n\in\mathcal{P}(\mathcal{X})$ with $n\in\mathbb{N}_+$, $(f(\Pi_1),\ldots,f(\Pi_n))^\intercal \sim\mathcal{N}(0, K)$, where $K_{ij}=\rho(\Pi_i, \Pi_j)$. To choose a suitable kernel, we can first define a kernel $k$ over $\mathcal{X}$ such that $k:\mathcal{X}\times\mathcal{X}\rightarrow\mathbb{R}$ and define the \textit{kernel mean embeddings} \citep{muandet2016kernel}
    \begin{talign*}
        \mu_P(\cdot):= \int_{\mathcal{X}} k(\cdot, x) \text{d}P(x)\in\mathcal{H}_k,
    \end{talign*}
    where $\mathcal{H}_k$ is the reproducing kernel Hilbert space (RKHS). With $\mathcal{H}_k$, we can now measure the similarity between distributions by the similarity between kernel mean embeddings via the inner product and norm of $\mathcal{H}_k$:
    \begin{talign}
        \label{eqn:rkhs}
        \langle \mu_P, \mu_Q\rangle_{\mathcal{H}_k}=&\int_{\mathcal{X}}\int_{\mathcal{X}} k(x,x') \text{d}P(x)\text{d}Q(x'),\\
         ||\mu_P - \mu_Q ||_{\mathcal{H}_k}^2=&  \langle \mu_P, \mu_P\rangle_{\mathcal{H}_k} + \langle \mu_Q, \mu_Q\rangle_{\mathcal{H}_k} - 2\langle \mu_P, \mu_Q\rangle_{\mathcal{H}_k} \nonumber,
    \end{talign}
    obtained using the kernel trick. This allows us to define a linear kernel $\rho(P, Q)=\langle \mu_P, \mu_Q\rangle_{\mathcal{H}_k}$ or a squared exponential kernel $\rho(P, Q)=\sigma^2\exp\left(-||\mu_P - \mu_Q ||_{\mathcal{H}_k}^2/(2\ell^2)\right)$, where $\sigma,\ell>0$. Lastly, given empirical measures $\hat{P},\hat{Q}$, we can estimate Equation~\ref{eqn:rkhs} with Monte Carlo integration 
    \begin{talign*}
        \langle \mu_P, \mu_Q\rangle_{\mathcal{H}_k}\approx\langle \mu_{\hat{P}}, \mu_{\hat{Q}}\rangle_{\mathcal{H}_k}=\frac{1}{N_i N_j}\sum_{l,m} k(x_{il},x_{jm}),
    \end{talign*}
    where $x_{il}\stackrel{iid}{\sim} P$ and $x_{jm}\stackrel{iid}{\sim} Q$.

    \paragraph{Aggregated Gaussian Processes:} A closely connected relative of Gaussian Processes over distributions are \textit{aggregated Gaussian processes} \citep{law_variational_2018, hamelijnck_multi-resolution_2019, tanaka_spatially_2019,yousefi_multi-task_2019,lucas_improving_2020, arambepola2020simulation}, where 
    \begin{talign}
    \label{eqn:bgp}
        y_i = \int_{\mathcal{X}} f(x) \text{d}\Pi_i(x) + \epsilon_i, \qquad \epsilon_i\sim\mathcal{N}(0, \sigma^2),
    \end{talign}
    where $f\sim\mathcal{GP}(m,k)$, $\Pi_i\in\mathcal{P}(\mathcal{X})$ and $\epsilon_i$ is an independently distributed Gaussian noise with $\sigma>0$. $m$ and $k$ are the mean and covariance functions (for time series $\mathcal{X}$, $k$ could be the signature kernel \citep{toth20a,lemercier_siggpde_2021}). The aggregated Gaussian processes construction has a very intuitive interpretation: the crop yield (measured in bushels per acre) for county $i$ is the average of the yields at the pixels containing croplands throughout the county. Using $\hat{\Pi}_i$, Equation~\ref{eqn:bgp} becomes
    \begin{talign}
    \label{eqn:model}
      y_i =  \mathbf{w}_i^\intercal \mathbf{f}_i + \epsilon_i, \qquad \epsilon_i\sim\mathcal{N}(0, \sigma^2),
    \end{talign}
    where $\mathbf{f}_i\defeq (f(x_{i1}),\ldots,f(x_{iN_i}))^\intercal$, and $\mathbf{w}_i \defeq  (w_{i1},\ldots,w_{iN_i})^\intercal$. Similarly, denote $\mathbf{f}_*\defeq (f_{*1},\ldots,f_{*N_*})^\intercal$ for prediction points $\{x_{*j}, w_{*j}\}_{j=1}^{N_*}$ with true label $y_*$. Suppose further that $\mathbf{X}_i \defeq  (x_{i1},\ldots,x_{iN_i})^\intercal$, $\mathbf{X}_*\defeq (x_{*1},\ldots,x_{*N_*})^\intercal$ and $k_{\mathbf{X}_i\mathbf{X}_j} := (k(x_{il}, x_{jm}))_{l,m}$ for pairs $i,j$, and $l=1,\ldots,N_i$ and $m=1,\ldots,N_j$,
    
    With $(K)_{ij} = (\mathbf{w}_i^\intercal k_{\mathbf{X}_i\mathbf{X}_j}\mathbf{w}_j)_{ij}$, $(K_*)_i:= (\mathbf{w}_i^\intercal k_{\mathbf{X}_i\mathbf{X}_*}\mathbf{w}_*)_i$ and $K_{**}:= \mathbf{w}_*^\intercal k_{\mathbf{X}_* \mathbf{X}_*}\mathbf{w}_*$, the posterior is $\mathbf{w}_*^\intercal \mathbf{f}_* | y \sim \mathcal{N}(\tilde{m}_{\mathbf{X}_*}, \tilde{k}_{\mathbf{X}_*\mathbf{X}_*} )$ with
    \begin{talign*}
        \tilde{m}_{\mathbf{X}_*} &\defeq \mathbf{w}_*^\intercal m_{\mathbf{X}_*} +  K_*^\intercal (K+\sigma^2 I_n)^{-1}(y - m_\mathbf{X}), \\ 
        \tilde{k}_{\mathbf{X}_*\mathbf{X}_*} &\defeq K_{**} - K_*^\intercal (K+\sigma^2 I_n)^{-1} K_*.
    \end{talign*}
    If elements of $\mathcal{P}(\mathcal{X})$ are known explicitly, we can replace the entries of $K,K_*$ and $K_{**}$ with exact integrals e.g. $K_{ij}=\int_{\mathcal{X}} \int_{\mathcal{X}} k(x,x') \text{d}P(u) \text{d}Q(v)$. Having inferred the posterior distribution of $f$ itself, it is then possible to directly query $f$ at the pixel level \citep{law_variational_2018}, or \textbf{disaggregate}, which can been used, e.g. for providing policy guidance in Malaria prevalence mapping \citep{lucas_improving_2020,arambepola2020simulation}. In addition, one can also query an unseen bag-level aggregate $\mathbf{w}_*^\intercal \mathbf{f}_*$.
    
    \paragraph{Connection to Gaussian Processes over Distributions:} We now establish the equivalence between aggregated Gaussian process and Gaussian processes over distributions with the kernel $\rho(P, Q)=\langle \mu_P, \mu_Q\rangle_{\mathcal{H}_k}$. If we consider a Gaussian process $g\sim\mathcal{GP}(0,\rho)$ with a distributional kernel $\rho$, then it can be shown that (\ref{eqn:bgp}) with $f\sim\mathcal{GP}(0,k)$, for any $\Pi_1,\ldots,\Pi_n\in\mathcal{P}(\mathcal{X})$, 
    \begin{talign*}
    (\int f \text{d}\Pi_1,\ldots,\int f \text{d}\Pi_n)^\intercal,\quad \text{and}\quad (g(\Pi_1),\ldots,g(\Pi_n))^\intercal
    \end{talign*}
    are equal in distribution, since $\rho(P, Q)=\int\int k(x,x') \text{d}P(x) \text{d}Q(x')$ for any $P,Q\in\mathcal{P}(\mathcal{X})$. However, this equivalence breaks down if we consider a potentially nonlinear $\rho$. In addition, despite the 2 models being equivalent, we cannot simply establish that $g(P)=\langle f,\mu_P\rangle_{\mathcal{H}_k} = \int f \text{d}P$ since $f$ lies outside of $\mathcal{H}_k$ almost surely \citep{kanagawa_gaussian_2018}. A major disadvantage of both Gaussian processes is the need to perform $\mathcal{O}(\sum_{i,j=1}^n N_i N_j)$ number operations to compute the matrix $K$, and a subsequent $\mathcal{O}(n^3)$ to invert $K+\sigma^2 I_n$, therefore prohibiting the ability to efficiently tune the hyperparameters via maximum likelihood or sample via Markov chain Monte Carlo.

    \paragraph{Variational Aggregated Gaussian Processes:}
    A major advantage for aggregated Gaussian processes, compared to Gaussian processes over distributions, is that the using a sparse variational Gaussian process approximation (SVGP; \citet{hensman_gaussian_2013,hensman_scalable_2015}) on aggregated Gaussian processes allows us to overcome the main computational bottleneck of distribution regression methods. \textbf{VBAgg} \citep{law_variational_2018, yousefi_multi-task_2019} proposes that we approximate $f$ directly with SVGP, making the approximation more feasible and allowing us to learn model hyperparameters via optimisation methods in mini-batches \citep{salimbeni2018natural, adam2021dual} in $\mathcal{O}(M^3 + M^2\sum_{i=1}^{n_b}N_i)$ for a batch size of $n_b$ and $M$ inducing points. Although we could also use the viewpoint from $g$ to formulate a sparse variational approximation of $g$, it is not clear how to select and learn the inducing points, which lie in $\mathcal{P}(\mathcal{X})$. Even if we were able to do so, we would also need multiple \textit{inducing distributions}, potentially still making the variational approximation very expensive. In addition, it is even more difficult as we only have access to empirical versions of elements in $\mathcal{P}(\mathcal{X})$.

    \paragraph{Data Preprocessing for Earth Observation Data:}
    Figure~\ref{fig:covariates_layered} illustrates the spatial resolutions of the different covariates of interest at De Witt County, Illinois, United States. We can see that each resolution contains pixels of varying sizes and different numbers of observations at different spatial locations, which was previously not addressed in past works for aggregated Gaussian processes. \textbf{Note:} each resolution $l$ may also contain multiple different covariates (with dimension equalling $D_l$). To compute the distributional kernel in Equation~\ref{eqn:rkhs}, we have to sample from $\Pi_i$, which is the joint distribution over all covariates and resolutions, requiring \textbf{resolution matching} to sample from it.
    
\begin{figure}[t]
    \centering
    \subfigure[\textbf{Data-Rep}]{
    \includegraphics[width=0.47\columnwidth]{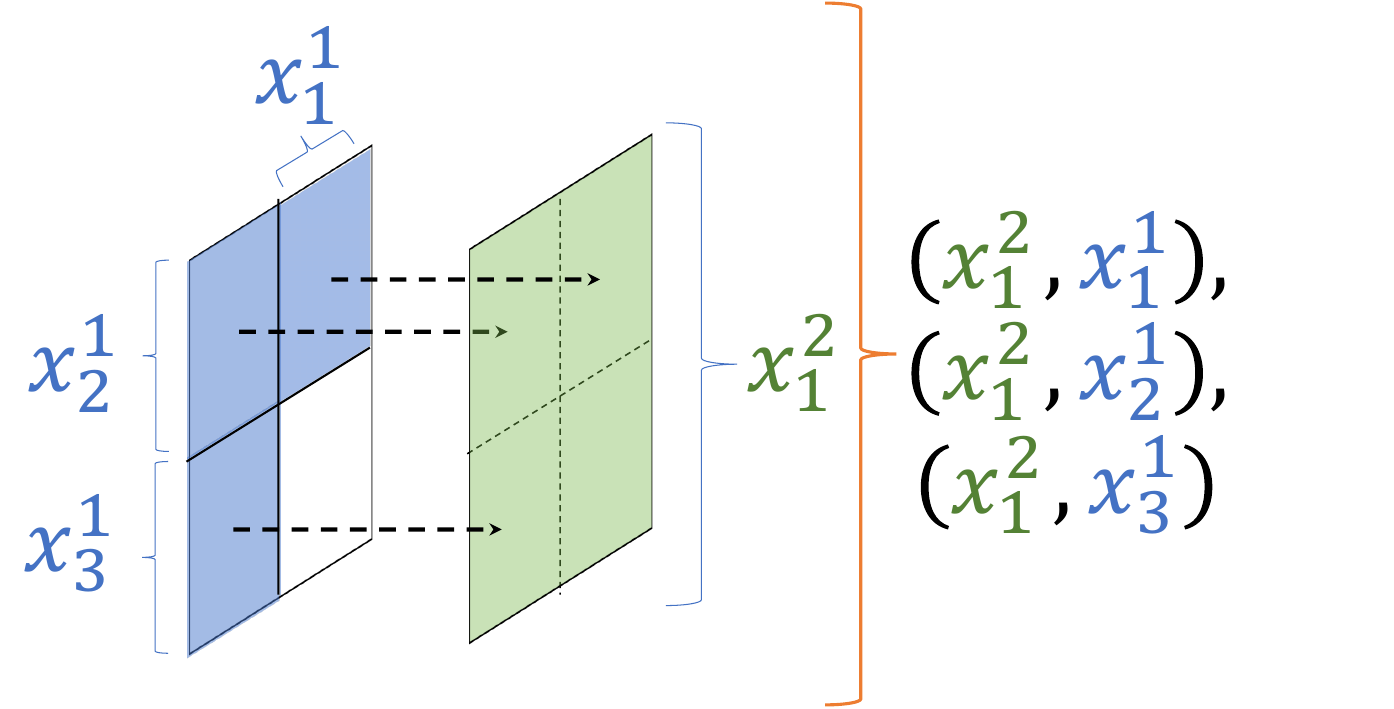}
    }
    \subfigure[\textbf{Data-Agg}]{
    \includegraphics[width=0.47\columnwidth]{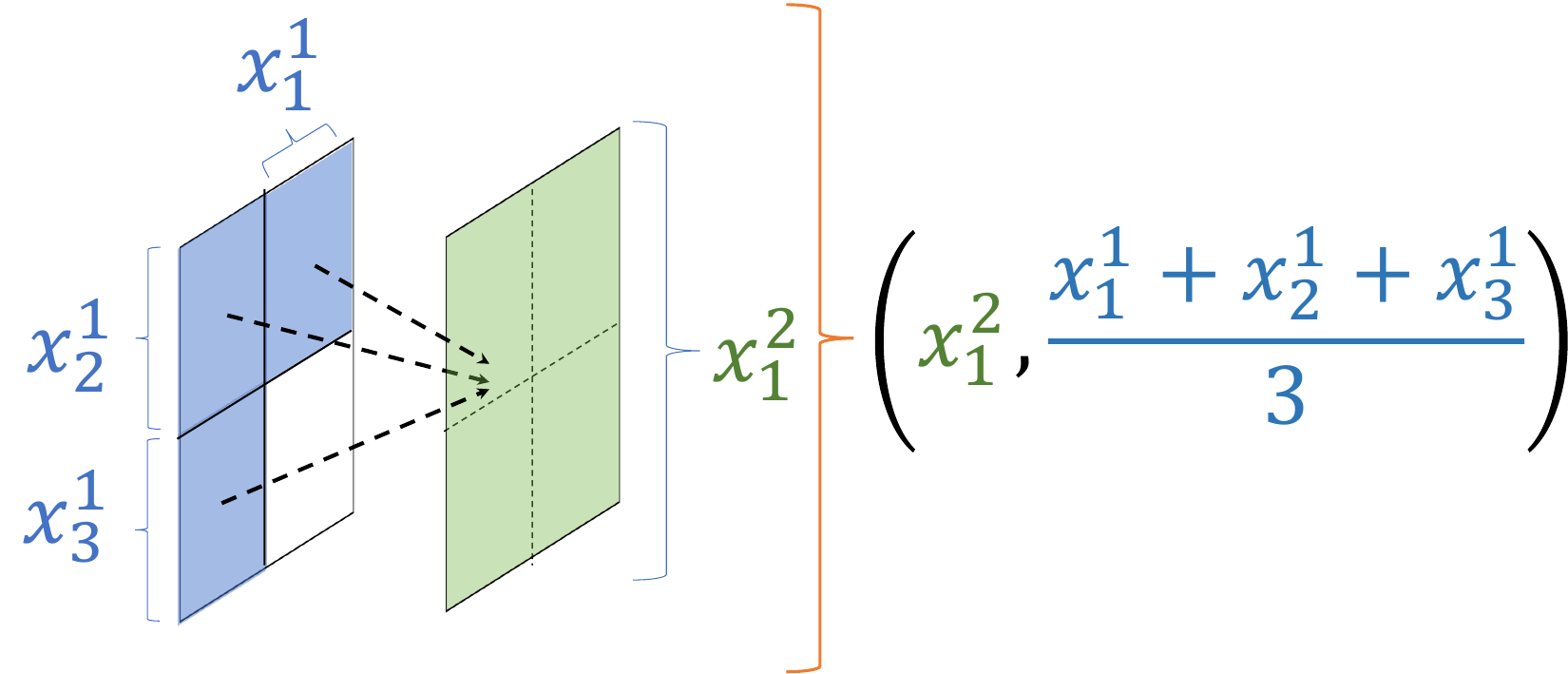}
    }
    \caption{Data preprocessing methods for Earth observation data.}
    \label{fig:data_preprocessing}
    \end{figure}

    Solutions that practitioners typically use (see Figure~\ref{fig:data_preprocessing}) are (1) \textbf{Data-Agg}: aggregating the covariates to a single resolution, typically the lowest resolution; (2) \textbf{Data-Rep}: repeat the pixels for the lower resolutions and obtain a potentially very large bag, if pixels at all resolutions are nested at each spatial location. To do this, we note that a pixel in the highest resolution lies in a nested set of lower resolution pixels from the other covariates, with each pixel having a single value associated with it. \textbf{Data-Agg} is commonly used and may result in loss of distributional information due to the aggregation, which we demonstrate in our experiments. \textbf{Data-Rep} may create large bags that are computationally infeasible, or if we subsample the data there may again be excessive losses of distributional information.

    \section{Multiresolution Aggregated Gaussian Processes}
    We propose \textbf{Multiresolution Aggregated Gaussian Processes} (MAgg) by exploiting the structure of additive kernels \citep{duvenaud2011additive} within aggregated Gaussian processes. We show that our model better handles the multiresolution nature of Earth observation data by minimising the amount of data preprocessing needed, whilst retaining the computational advantages of sparse variational aggregated Gaussian processes and gaining additional interpretability.
    
    Let $\mathcal{X}=\mathcal{X}^1\times\cdots\times\mathcal{X}^d$ and $\Pi$ have marginal distributions $\Pi^1\ldots\Pi^d$. The constant $d$ denotes the number of different resolutions, in which each itself may also contain different covariates. Treating each bag of covariates as a distribution, for each response $y_i$ we have the marginal distributions $\Pi^1_i\in\mathcal{P}(\mathcal{X}^1),\ldots,\Pi^d_i\in\mathcal{P}(\mathcal{X}^d)$. We integrate an additive Gaussian process $f$
    \begin{talign*}
    y_i &= \int_{\mathcal{X}} f(x) \text{d}\Pi_i(x) +\epsilon_i\\
    &=\sum_{l=1}^d \int_{\mathcal{X}^l} f^l(x^l) \text{d}\Pi_i^l(x^l) + \epsilon_i, \epsilon_i\sim\mathcal{N}(0,\sigma^2),
    \end{talign*}
    where $f\sim\mathcal{GP}(0, \sum_{l}k_l)$ and $f^l\sim\mathcal{GP}(0, k_l)$ with $k^l:\mathcal{X}^l\times\mathcal{X}^l\rightarrow\mathbb{R}$. Each Gaussian process $f^l$ only depends on the $l$th resolution, meaning that we only need to consider integrating over the marginal distribution $\Pi^l$, enabling us to circumvent any resolution matching.
    
    We may also model the interaction between different covariates at different resolutions using the additive kernel of \citet{duvenaud2011additive} and thus the corresponding aggregated model 
    \begin{talign*}
      y_i =&  \int f^1 \text{d}\Pi_i^1+\cdots+\int f^{12}\text{d}\Pi_i^{12}+\cdots +\int f^{1\cdots d} \text{d}\Pi_i^{1\cdots d}+ \epsilon_i,\qquad \epsilon_i\sim\mathcal{N}(0,\sigma^2),
    \end{talign*}
    where $f\sim\mathcal{GP}(0, \sigma_1^2\sum_{l=1}^d k^l + \sigma_2^2\sum_{l,m: l\neq m}k^l k^m +\cdots+\sigma_d^2\prod_{l=1}^d k^l)$, $\sigma_1\ldots,\sigma_d>0$ are scale factors for each order of interaction and $\Pi^u_i$ is the joint distribution over a set of resolutions $u$ for $u\subseteq[d]=\{1,\ldots,d\}$. 
    
    Denote the inner product distributional kernel in Equation~\ref{eqn:rkhs} as $\bar{\Pi}^u_i\Pi^u_j[k^u]:= \int \int \prod_{l\in u} k^l(x,x') \text{d}\Pi^u_i(x)\text{d}\Pi^u_j(x')$, where $u\subseteq [d]$. To compute the variance of $\int f\text{d}\Pi_i$, which reduces to computing $\bar{\Pi}_i\Pi_j[k]$, one can decompose the integral into 
    \begin{talign*}
        \bar{\Pi}_i\Pi_j[k] &= \sigma_1^2\sum_{l=1}^d \bar{\Pi}_i^l\Pi_j^l[k^l]  +\cdots+\sigma_d^2 \bar{\Pi}_i^{1\cdots d}\Pi^{1\cdots d}_j[\prod_{l=1}^d k^l],
    \end{talign*}
    where we can compute each term in the total summation with the empirical distribution $\hat{\Pi}^u_i$ for the subset of resolutions $u$. The advantage of this decomposition is that for each $\hat{\Pi}^u_i$, resolution-matching (see Section 2) \textbf{only needs to occur between resolutions} in $u$. For $\Pi_i^{1\cdots d}$, the joint distribution over all resolutions, we match all resolutions as for VBAgg. However, to compute the first order effects we do not require any preprocessing at all, leaving us with the original $N_{il}$ number of pixels. For second order effects, given resolutions $l$ and $h$, we only need to resolution-match the $N_{il}$ and $N_{ih}$ number of pixels from each resolution.

    \paragraph{Variational Inference:}  We propose \textbf{MVBAgg}, a SVGP approximation to our model. Although we have constructed an additive kernel that aggregates each component separately, we still need to address the computational bottleneck of aggregated Gaussian processes with SVGP \citep{hensman_gaussian_2013,hensman_scalable_2015}. Since the sample path is $f=f^1+\cdots+f^{1\ldots d}:\mathcal{X}\rightarrow\mathbb{R}$, this suggest that we can still define the usual inducing variables by selecting $M$ inducing points $\Z=(z_1,\ldots,z_M)^\intercal\subset \mathcal{X}$, and setting the inducing variable to be $\mathbf{u}=f(\Z)\equiv(f(z_1),\ldots,f(z_M))^\intercal$. Similar to \citet{law_variational_2018}, we select $\Z$ by first applying \textbf{Data-Rep} to each training bag and then picking 1 cluster centre selected via KMeans. This step is only done outside the training loop to initialise the inducing points, so that we can take into account of correlations between resolutions and preserve distributional information, and is feasible as KMeans is very scalable. One may also pick a cluster centre for each resolution and then concatenate to obtain the subsequent inducing point for a bag, yielding similar $\Z$'s but losing some inter-resolution correlations.

    
    We pose a variational distribution $q(f,\mathbf{u})\defeq  p(f|\mathbf{u})q(\mathbf{u})$ and construct a variational approximation $q(\mathbf{u})= \text{argmin}_{q\in\mathcal{Q}} \text{KL}(q(f,\mathbf{u})||p(f,\mathbf{u}|y))$. We choose the variational family $\mathcal{Q}=\{q(\mathbf{u})=N(\mathbf{u};\m, \S): \m\in\mathbb{R}^{M}, \S\in\mathbb{R}^{M\times M}\}$. We thus have the approximate posterior $q(f)\defeq \int p(f|\mathbf{u})q(\mathbf{u}) \text{d}\mathbf{u}\equiv \mathcal{GP}(f;\tilde{m}, \tilde{k})$  
    where
    \begin{talign*}
        \tilde{m}(x) = k_{x\Z}\alpha,\quad 
        \tilde{k}(x,x') = k_{xx'} - k_{x\Z}\mathbf{Q}^{-1}k_{\Z x'},
    \end{talign*}
    where $\alpha=k_{\Z\Z}^{-1}\m$ and $\mathbf{Q}^{-1}=(k_{\Z\Z}^{-1} - k_{\Z\Z}^{-1}\S k_{\Z\Z}^{-1})$. Aggregating gives $\int f \text{d}\Pi_i\sim\mathcal{N}\bigg(\Pi_i[\tilde{m}],  \bar{\Pi}_i\Pi_i[\tilde{k}]\bigg)$,
    where $\Pi_i[\tilde{m}] =\int \tilde{m}(x) \text{d}\Pi_i(x)$. Denoting 
    \begin{talign*}
    \Pi_i^u[k^u_{\cdot\Z}]&:= \int \prod_{l\in u}k^l_{x\Z}\text{d}\Pi^u_i(x),\quad \Pi_i^u[k^u_{\Z\cdot}]:= \Pi_i^u[k^u_{\cdot\Z}]^\intercal,
    \end{talign*}
    where $u\subseteq[d]$, we have 
    \begin{talign*}
        \Pi_i[k_{\cdot\Z}] &= \sigma_1^2\sum_{l=1}^d \Pi_i^l[k^l_{\cdot \Z}]  +\cdots+\sigma_d^2\Pi^{1\cdots d}_i[\prod_{l=1}^d k^l_{\cdot \Z}],\\
        \bar{\Pi}_i\Pi_i[k] &= \sigma_1^2\sum_{l=1}^d \bar{\Pi}^l_i\Pi^l_i[k^l]+\cdots+\sigma_d^2 \bar{\Pi}_i^{1\cdots d}\Pi_i^{1\cdots d}[\prod_{l=1}^d k^l],
    \end{talign*}
    giving
    \begin{talign*}
        \Pi_i[\tilde{m}] &=\Pi_i[k_{\cdot\Z}]\alpha,\\
        \bar{\Pi}_i\Pi_i[\tilde{k}] &= \bar{\Pi}_i\Pi_i[k] - \Pi_i[k_{\cdot\Z}]\mathbf{Q}^{-1}\Pi_i[k_{\Z \cdot}].
    \end{talign*}
    We see that each pair of $\Pi^u_i[k^u_{\cdot\Z}]$ and $\bar{\Pi}^u_i\Pi^u_i[k^u]$ can be aggregated with respect to its own resolution using $\Pi_i^u$ again. With the empirical measures $\hat{\Pi}^u_i$, we have the approximations $\Pi^u_i[k^u_{\cdot\Z}]\approx (\mathbf{w}_i^u)^\intercal k_{\mathbf{X}_i^u\Z}$ and $\bar{\Pi}_i^u\Pi^u_i[k^u]\approx (\mathbf{w}_i^u)^\intercal k_{\mathbf{X}_i^u\mathbf{X}_i^u}\mathbf{w}_i^u$. Note that we cannot apply the Newton-Girard rule for computing additive kernels \citep{duvenaud2011additive} since the distributions $\Pi^u_i$ for $u\subseteq[d]$ are not a product distributions in general (i.e. independence between resolutions). However, for practical applications such as crop yield modelling, it may be sufficient to use only 3-4 data sources (each with a different resolution) and limit the order of interactions to 2, giving us $d + d(d-1)/2$ terms to compute. Indeed as discussed in \citet{duvenaud2011additive}, many problems may not require very high orders of interactions. 
    
    Lastly, the hyperparameters, such as the kernel parameters, can be learned via maximisation of the lower bound
    \begin{talign*}
    \mathcal{L}\defeq  \text{KL}(q(\mathbf{u}) || p(\mathbf{u})) - \sum_{i=1}^n \mathbb{E}_{q(f)}[\log p(y_i| \int f \text{d}\Pi_i)],
    \end{talign*}
    and solved via optimisation methods in mini-batches \citep{salimbeni2018natural, adam2021dual} again. We note that non-Gaussian likelihoods $p(y_i| \int f \text{d}\Pi_i)$ are also possible, such as Poisson \citep{law_bayesian_2018} or Binomial. Inheriting the nice properties of VBAgg, with order 1 interactions for instance, we can attain $\mathcal{O}(M^3 + M^2 \sum_{l=1}^d\sum_{i=1}^{n_b} N_{il})$ complexity per iteration.

    \paragraph{Interpretability:} We can \textbf{disaggregate} or make pixel-level predictions for each component $f^1,\ldots,f^{12},\ldots,f^{1\cdots d}$, which are functions at the covariate resolution. Despite the non-identifiability of each function (our estimates of each function will always have a constant bias), we may nonetheless interpret the gradients of the disaggregated maps \citep{law_variational_2018,lucas_improving_2020, agarwal2020neural} for decision-making. Post-training, the multiresolution viewpoint allows disaggregation of $f$ at the highest resolution if we use \textbf{Data-Rep} to obtain a single vector $x$ for all the covariates at the highest resolution, for which we can use to calculate the posterior distribution of $f(x)$. In comparison, VBAgg with \textbf{Data-Agg} only has 1 resolution to disaggregate to because all the resolutions have already been matched.
    
    Another way to gain interpretability from aggregated Gaussian processes is to determine the covariate sensitivity. For simplicity, only considering 1st order interactions, we can calculate the 1st and 2nd order Sobol indices \citep{sobol1990sensitivity}:
    \begin{talign*}
        S_l = \frac{\text{Var}_{x}[{\tilde{m}^l(x_l)}]}{\text{Var}_x[\tilde{m}(x)]+\sigma^2},
        S_{u} = \frac{2\text{Cov}_{x_l, x_h}[\tilde{m}^l(x_l), \tilde{m}^h(x_h)]}{\text{Var}_x[\tilde{m}(x)]+\sigma^2},
    \end{talign*}
    where $u=\{l,h\}$ is a tuple and we approximated each Sobol index term with Monte Carlo using the entire dataset obtained with \textbf{Data-Rep} for MVBAgg. The 1st order Sobol indices indicate how much the variance is explained by each resolution on its own, and the 2nd order Sobol indices between pairs of resolutions. We note that $\sum_{l=1}^d S_l +\sum_{u:u=(l,h)} S_u + \sigma^2=1$ and it also possible to consider higher order Sobol indices to take into account of more interactions.  Additionally, we can also compute the Sobol indices for each distribution $\Pi_i$ to obtain local sensitivity analysis within each bag, for which the Var and Cov operators are integrated with respect to $\Pi_i$. More details on the approximate of Sobol indices is in Appendix~\ref{appendix:experiments}. 
    
    \section{Related Work}
    
    \paragraph{Aggregated Gaussian Processes:} In previous aggregated Gaussian process works, all the resolutions are usually preprocessed to the same resolution \citep{law_variational_2018,tanaka_spatially_2019,yousefi_multi-task_2019,hamelijnck_multi-resolution_2019, arambepola2020simulation,lucas_improving_2020}. As mentioned at the start of Section 3, this may have disadvantages. In contrast, our model only requires full resolution matching if we use the highest order of interaction within the additive kernel. In the context of classification of images, alternative aggregation methods are max-aggregation or attention-aggregation \citep{kim_gaussian_2010, haussmann_variational_2017,  ilse_attention-based_2018}, which are special cases of $\Pi_i=\delta(\max_{x\in N_i^\complement})$, where $N_i=\bigcup_{N\in\mathcal{F}} N$ such that $\Pi_i(N)=0$, the union of all sets in $\mathcal{X}$ with zero measure under $\Pi_i$. In the spatial statistics literature \citep{gelfand_handbook_2010,diggle2013spatial, wilson2020pointless}, what are referred to as spatially aggregated or Gaussian processes over areal data are also aggregated Gaussian processes.

    \paragraph{Disaggregation:} Deconditional mean embeddings \citep{chau2021deconditional} and 2-staged ridge regression \citep{stefanovic2021reconstructing} have recently used for disaggregation, although both methods require mediating variables at the same resolution as the response, which is less related to the challenges that we address in this paper. Additionally, \citet{law_bayesian_2018,hamelijnck_multi-resolution_2019,lucas_improving_2020, arambepola2020simulation} all address the problem of disaggregation for the purpose of public policy guidance but only for single, preprocessed resolutions obtain with \textbf{Data-Agg}, whereas our model is able to disaggregate at the highest resolution and better maintain fine-scale covariate information.
    
    \paragraph{Multiresolution Regression:} Multiresolution Gaussian processes \citep{fox2012multiresolution,pmlr-v89-taghia19a} and kernels \citep{cuturi2005multiresolution} assume that there is a nested partition of $\mathcal{X}$. This setting would be equivalent to applying \textbf{Data-Agg} to match all resolutions to one, if there exists a nested structure. In our case, we do not assume that there is always such a structure, which helps us increase the number of Monte Carlo points we can use for lower order interactions. Furthermore, \citet{rudner2018rapid} developed $\text{Multi}^3\text{net}$ to also handle the multi-resolution nature of different satellite imageries for regression. In their work, they used an ensemble of independent neural networks to learn from multiresolution images, whereas we consider more parameter efficient models through kernels and learn from set or distribution-valued inputs.
    
    Our work is also closely related to multi-source distribution regression of \citet{thorns_distribution_2018,adsuara_nonlinear_2019}, where the aggregations are also done over separate resolutions. In their works, they consider distribution regression fitted via random Fourier features \citep{rahimi_random_2008} and Bayesian distribution regression \citep{law_bayesian_2018}, which yield similar computational complexities as MVBAgg and are both applied to crop yield prediction. However, both works only consider bag-level predictions, whereas we also study disaggregation.
    
    \paragraph{Distribution Regression:} Closely related to Gaussian processes over distributions, distribution regression \citep{szabo_learning_2016} also provides a family of frequentist regression methods to learn from distributions. A regression function $f\in\mathcal{H}_\rho$ is learned via regularised least squares over the dataset $\{(\mu_{\hat{\Pi}_i},y_i)\}_{i=1}^n$. Distribution regression methods have been successfully applied to crop yield prediction \citep{thorns_distribution_2018,adsuara_nonlinear_2019, lemercier_distribution_2021}, election forecasting \citep{flaxman2016understanding} and sequential data \citep{lemercier_distribution_2021}. Bayesian approaches can also be used to model within-bag variations of the covariates for downstream regression tasks \citep{law_bayesian_2018}. To note, although not done explicitly for existing works, for both Gaussian processes over distributions and distribution regression, it is also possible to make `individual level' predictions at the Dirac measures $\delta_{x_{ij}}$.
    
    \paragraph{Crop Yield Modelling:} Our main experiment models crop yields using climatic and satellite data, which are also widely used in previous works \citep{burke_using_2021}. Many previous works have either spatially averaged the pixels for each county \citep{mateo-sanchis_synergistic_2019,sanchis_multisensor_2019,martinez2020crop,han2020prediction, fan2021a} or summarised the spatially-distributed pixels into an empirical distribution (histogram) \citep{you_deep_2017}. \citet{you_deep_2017} assumes permutation invariance of the pixels, which we also assume, but our method would retain more distributional information as we implicitly use empirical kernel mean embeddings. \citet{brus2018geostatistical} also studies disaggregation for crop yields using aggregated Gaussian processes, but models all other covariates other than longitude and latitude using a linear model.
    
    \section{Experiments}
    We demonstrate the predictive performance of MVBAgg compared to VBAgg, Random Forest and a Gaussian process regression model (CGP) with centroid covariates. For VBAgg, we apply \textbf{Data-Agg} via mean aggregation to the lon-lat and MODIS covariates match them to GRIDMET resolution. As noted in Table~\ref{tbl:data}, we apply \textbf{Data-Agg} to the lon-lat and MODIS covariates to slightly upsample them from 30m to 500m and 250 to 1000m resolutions respectively, due to the sheer size of the data. For simplicity, we ignore the temporal resolution in the experiments by concatenating the temporal resolution at each spatial location to work with a 13-dimensional vector. The time window taken is between April and October, which was noted in \citet{mateo-sanchis_synergistic_2019} as the crop growth window. Since the MODIS covariates are measured every 13 days, we take samples for GRIDMET covariates on the same days as well. Lastly, we do not apply a crop mask for lon-lat due to the fact that the distributions are uniform distributions (i.e. measures of area).
    
    When training VBAgg and MVBAgg, we perform natural gradient descent \citep{salimbeni2018natural} to learn $q(\mathbf{u})$ and learn all other hyperparameters with the Adam optimiser. Details of each model and configurations are available in Appendix~\ref{appendix:experiments}. We use 1st order interactions for simplicity. For computational tractability during training, we restrict the maximum bag size to $N_i=100$ for MODIS and GRIDMET covariates, and $N_i=500$ for lon-lat, via random subsampling with each bag. This is because for MODIS, the bag sizes can range from 67 to 5950, whereas for lon-lat the range is between $2399$ to $34867$. We evaluate the predictive performance in estimating $y_i$ via the test loglikelihood (LL; higher is better) and root mean squared error (RMSE; lower is better). Our implementation (see Appendix~\ref{appendix:implementation}) relies on GPFlow \citep{de_g_matthews_gpflow_2017}, and all data and code will be made available upon publication. 
    
    \begin{table}[t]
        \caption{Collected datasets and their associated descriptions. We slightly downsampled the EVI from 250m to 1000m and latitude, longitude from 30m to 500m due to computational constraints in obtaining the dataset.}
        \begin{adjustbox}{width=\columnwidth,center}
        \begin{tabular}[t]{| c | c | c| }
        \hline
        Data  & Source & Resolution  \\
        \hline
        \hline
        Soybean mask & USDA NASS Cropmask \citep{nass2016usda} & 30m  \\
        latitude, longitude ($D_1=2$) & USDA NASS Cropmask \citep{nass2016usda} & 500m  \\
        EVI ($D_2=13$) & MODIS \citep{didan_mod13q1_2015} & 1000m  \\
        pr, tmmx ($D_3=26$) & GRIDMET \citep{gridmet}  &  4638.3m\\
        Soybean yield & USDA National Agricultural Statistics Service. \citep{statsusda} &  US County\\
        \hline
        \end{tabular}
        \end{adjustbox}
        \label{tbl:data}
    \end{table}

    \begin{table}[t]
        \caption{Predictive performance over 5 different randomly selected set of counties. We report the average test LL and RMSE, alongside their standard errors. }
        \begin{adjustbox}{width=0.8\columnwidth,center}
        \begin{tabular}[t]{| c | c | c| }
        \hline
        Method  & LL & RMSE  \\
        \hline
        \hline
        Random Forest & NA &  3.236$\pm0.5975$\\
        CGP & 0.5779$\pm0.09628$ &  0.9458$\pm0.06792$\\
        VBAgg & 0.6976$\pm0.1771$ &  0.9371$\pm0.02296$\\
        MVBAgg & $\mathbf{0.9841\pm0.1046}$&  $\mathbf{0.6899\pm0.1185}$\\
        \hline
        \end{tabular}
        \end{adjustbox}
        \label{tbl:synthetic_results}
    \end{table}
    
    \subsection{Semi-synthetic} 
    We generate synthetic aggregated labels using real covariates from satellite imagery and weather features over croplands collected from Google Earth Engine \citep{gorelick2017google}. We include the covariates: latitude, longitude, enhanced vegetation index (EVI) on 7th April 2015, precipitation (pr) and maximum temperature (tmmx) on 7th April 2015. For data generation, we use the full bag and generate
    \begin{talign*}
    y_i =& \sum_{l=1}^4  \frac{1}{N_{il}}\sum_{j=1}^{N_{il}} f^l(x_{ij}^l) + \epsilon_i,\quad \epsilon_i\sim\mathcal{N}(0,0.5^2).
    \end{talign*}
    £We let $f^0(x)=0.5(x_\text{lon}^2+x_\text{lat}^2)$, $f^1(x)=x_\text{EVI}^2$, $f^2(x)=1.5x_\text{pr}^2$ and $f^3(x)=2x_\text{tmmx}^2$.
    
    We report statistically signficant predictive results for MVBAgg in Table~\ref{tbl:synthetic_results}. A priori, we would expect CGP and Random Forest to perform worse as it only contains information about the first moments of each covariate distribution. In addition, we would not expect VBAgg to perform as well as MVBAgg due to (1) the preprocessing of the covariates to the GRIDMET resolution causes the distributions for latitude, longitude and EVI to be different from the underlying true aggregation distributions and (2) fewer points are used for Monte Carlo estimation of $\Pi_i[k_{\cdot\Z}]$ and $\bar{\Pi}_i\Pi_i[k]$. 
    
\begin{figure}[t]
        \centering
        \includegraphics[width = 0.5\columnwidth]{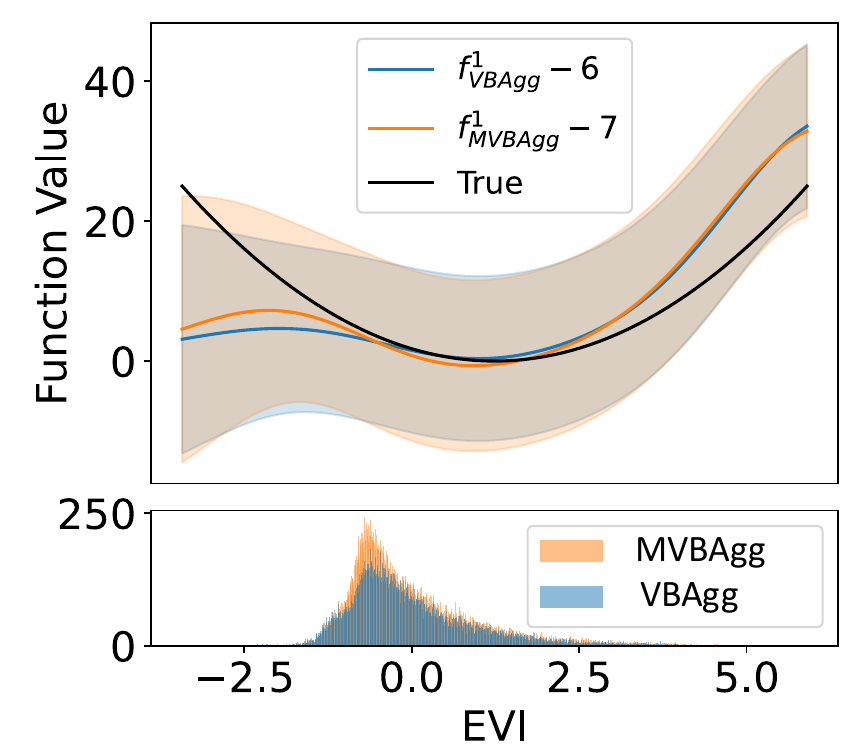}
        \caption{Disaggregation of $f^1$ (EVI covariates) via VBAgg and MVBAgg with 95\% credible intervals. The histogram shows the EVIs values used for each model.}
        \label{fig:synthetic_disagg}
    \end{figure}

    In addition, we can also disaggregate MVBAgg and VBAgg to obtain the underlying functions $f^l$, as shown in Figure~\ref{fig:synthetic_disagg}. Theoretically, it is expected that there are constant biases for each estimated $f^l$ because if the constant biases for each function cancel each other out, then they are also valid estimates for $f$. We therefore manually apply shifts to illustrate that we have indeed fitted each $f^l$ up to a constant shift. Although only having biased estimates of $f^l$, we can still interpret the gradient $f^l$ to make policy decisions (e.g. increases in EVI are associated with increases with yield) and $\text{Var}_x[f^l(x)]$ (e.g. to compute feature sensitivity). We see that the disaggregations for VBAgg and MVBAgg are very similar. In addition, the histogram of the EVI from all the bags shows that there are more values lying in $[-1,0]$ in the original data compared to the aggregated data that VBAgg uses, and it coincides with the fact that the disaggregation within that interval is more accurate.

    \subsection{Crop Yield Modelling} 
\begin{figure}[t]
        \centering
        \includegraphics[width = 0.7\columnwidth]{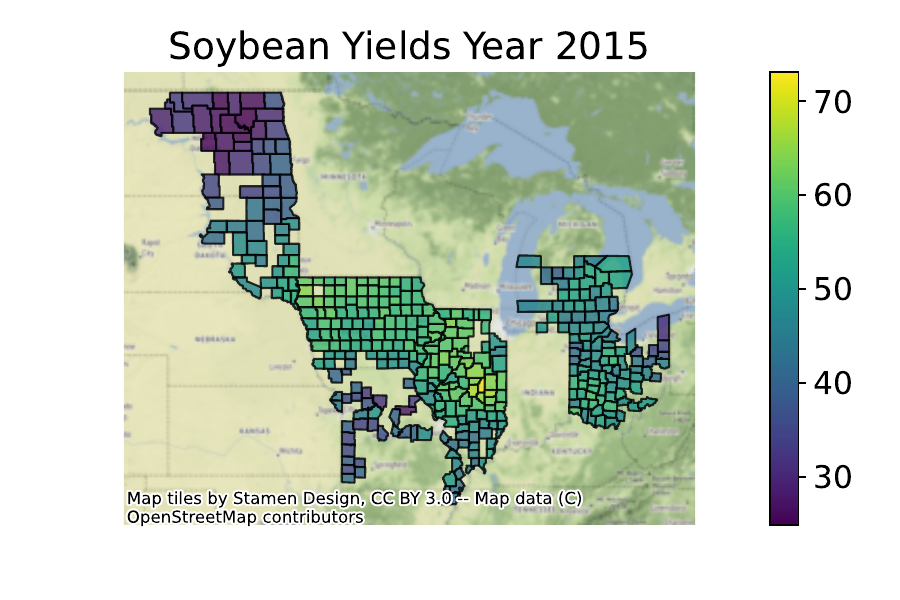}
        \caption{Soybean yields (bushels/acre) in the United States.}
        \label{fig:soybean_yields}
    \end{figure}

    By definition, crop yield is the average productivity/crop production per unit, which is in this case bushels/acre. Understanding crop yields can allow us to infer the fertility of the land as a function of climatic and satellite observations, allowing us to decide where to plant crops or which areas require policy interventions. We collected soybean yields (bushels/acre) from 2015 and 2017 for 384 counties in the crop belt of the United States (see Figure~\ref{fig:soybean_yields}). We use the same covariates previously used for the semi-synthetic data and work with log-yields in order to remove any skew.

    We randomly select sets of counties, train the models over 2015 data ($n=308$) and then make predictions for 2017 yields ($n=308$). In Table~\ref{tbl:crops_results} and \ref{tbl:crops_sobol}, we report the predictive performance and covariate sensitivities respectively. We see that MVBAgg again performs the best in terms of LL. The quality of bag-level predictions is in this case similar to CGP. We suggest that this is due to (1) the aggregation steps used for obtaining MODIS and lon-lat covariates due to computational constraints; (2) lack of additional informative covariates or (3) the crop yields only depend on the first moments of the covariates. Interestingly, we see that VBAgg performs a lot worse than CGP and MVBAgg, having aggregated both EVI and lon-lat covariates to GRIDMET's resolution. Therefore there is evidence to suggest that the data preprocessing step to match resolutions of covariates actually has a detrimental effect on the model performance of VBAgg, even though VBAgg uses a more sophisticated aggregation model than CGP.

    \begin{figure}[t]
        \centering
        \includegraphics[width = 0.47\columnwidth]{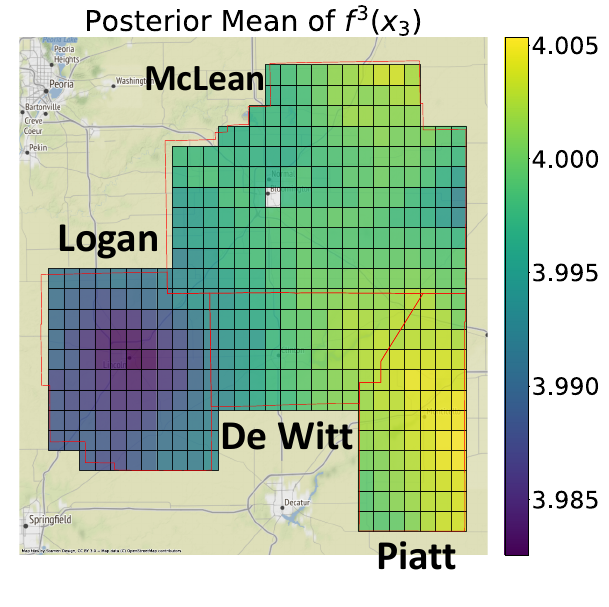}
        \includegraphics[width = 0.47\columnwidth]{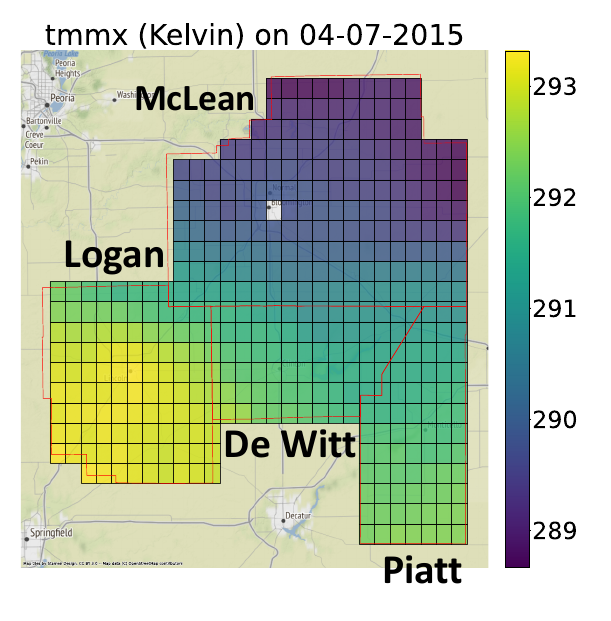}
        \caption{Disaggregation of $f^3$ (all 13 tmmx covariates) and observation of tmmx on the 7th of April, 2015. No disaggregations for Bloomington due to it being filtered out with the crop mask.}
        \label{fig:tmmx_disagg}
    \end{figure}
    
    \begin{figure}[t]
        \centering
        \includegraphics[width = 0.6\columnwidth]{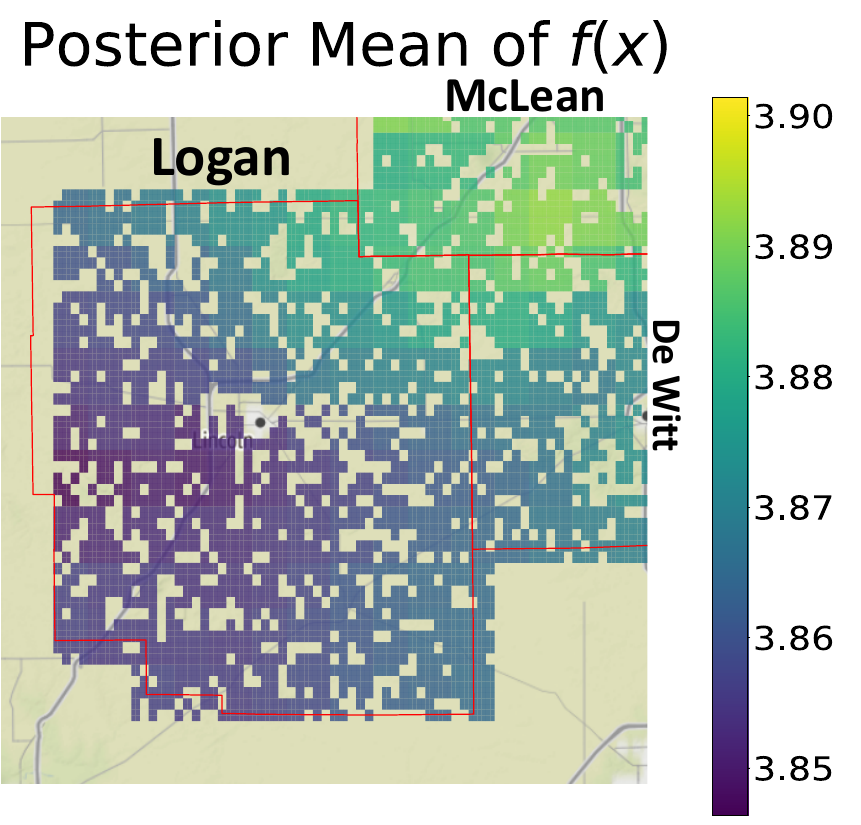}
        \caption{Disaggregation of $f$ (all covariates) and observation of tmmx on the 7th of April, 2015. No disaggregations at cities and rivers due to them being filtered out with the crop mask.}
        \label{fig:crop_disagg}
    \end{figure}
    
    From Table~\ref{tbl:crops_sobol}, we also see that CGP explains the variance in the response differently to MVBAgg, but we note that CGP uses only the centroids as input and thus it only explains how the covariate means affect the response. In addition, the likelihood noise almost accounts for no variability, which may also not be a realistic assumption as in practice crop yields can have a lot of noise. On the other hand, VBAgg identifies a similar set of covariates, especially the EVI, and seems to attribute a lot of the variance due to spatial randomness (i.e. lonlat). The likelihood noise also accounts for a smaller percentage of variance compared to MVBAgg. Considering the poor predictive performance, this raises the suspicion that the resolution-matching has made VBAgg pick up incorrect signals from the aggregated EVI and overfitted on lon-lat. MVBAgg on the other hand lies between CGP and VBAgg, whilst being able to explain covariates at higher resolutions with \textbf{Data-Rep} (VBAgg is limited to the covariates obtained via \textbf{Data-Agg}).
    
    From Figure~\ref{fig:tmmx_disagg} and \ref{fig:crop_disagg}, we can see that the model assigns lower yields in Logan county, and the Sobol indices in Table~\ref{tbl:crops_sobol} suggest that a large proportion of this can be explained by tmmx. Plotting 1 day out of 13 days of tmmx in 2015, we can see that Logan county saw relatively high temperatures, going up to 293. Soybean crops may fail at excessively high temperatures and it is probable that Logan county had consistently higher temperatures throughout the year than the other counties that made soybean yields lower.
    \begin{table}[t]
        \caption{Predictive performance over 5 different randomly selected set of counties. We report the average test LL and RMSE, alongside their standard errors.}
        \begin{adjustbox}{width=0.8\columnwidth,center}
        \begin{tabular}[t]{| c | c | c| }
        \hline
        Method  & LL & RMSE  \\
        \hline
        \hline
        Random Forest & NA &  0.1071$\pm0.002171$\\
        CGP & -0.7633$\pm0.01650$ &  0.1027$\pm0.002082$\\
        VBAgg & -0.8482$\pm0.01059$ &  0.1139$\pm0.001149$\\
        MVBAgg & $\mathbf{-0.7387\pm0.01575}$&  $\mathbf{0.1023\pm0.002372}$\\
        \hline
        \end{tabular}
        \end{adjustbox}
        \label{tbl:crops_results}
    \end{table}
    \begin{table}[t]
    \caption{Top 3 Sobol indices and estimated normalised likelihood variances based on 2015 data.}
    \begin{adjustbox}{width=0.8\columnwidth,center}
        \begin{tabular}[t]{| c | c| c| }
            \hline
            CGP & VBAgg& MVBAgg  \\
            \hline
            \hline
            EVI (0.551)  & lonlat-EVI (0.251) &   tmmx (0.442)\\
            EVI-tmmx (0.332)  &  EVI (0.231)  & lonlat-tmmx (0.186)\\
            lonlat-EVI (0.0222)  &  lonlat (0.199) & pr-tmmx (0.0647)\\
            \hline
            \hline
            $\frac{\sigma^2}{\text{Var}_x[\tilde{m}(x)]+\sigma^2}$ \vline\vline \quad $7.06e-5$  &   0.0894 & 0.185\\
            \hline
        \end{tabular}
    \end{adjustbox}
    \label{tbl:crops_sobol}
    \end{table}


    \section{Conclusion and Discussion}
    Motivated by the problem of crop yield modelling, where a crop yield response is paired with samples of many covariates collected from data sources of varying spatial resolutions, we propose a Gaussian process model to efficiently handle such multiple resolutions through an additive kernel structure and variational inference. Compared to existing approaches, we show that our method minimises the amount of data preprocessing needed to match between resolutions and better maintains all available distributional information. Through synthetic and real world crop yield modelling experiments, we demonstrate that commonly used data preprocessing techniques of matching resolutions via data aggregation can have negative consequences on the predictive performance and interpretability of aggregated Gaussian processes, and that our method overcomes these issues.  We hope that the methodology and workflow developed in this work can be used for further scientific investigations into problems of similar settings, such as crop yield modelling, epidemiology and climate science.

    \paragraph{Limitations and Future Directions:}
    In the alternative formulation with Gaussian processes defined over distributions, the aggregated Gaussian processes is restricted to the inner product second-level kernel, making the induced function space theoretically less expressive than general distributional kernels. Other types of kernels, however, do not admit an aggregated formulation, are less interpretable and it is unclear how to apply useful variational approximations to them. It would also be interesting to explore more expressive formulations of aggregated Gaussian processes, e.g. alternative aggregation operators.
    
    In our experiments, for computational scalability, we subsampled each bag and set a limit on the number of items in each bag. However, given more computational resources, this may not be necessary and may yield better performances. We stress that the crop yield experiments presented here are only for demonstration purposes: we provide the modelling tools and workflow that may be able to generate impactful insights when combined with domain expertise, especially when deciding which type of satellite data and which covariates to use.

    \bibliography{ref}
    \bibliographystyle{plainnat}

    \newpage
    \onecolumn
    \appendix
    
    \section{Derivation of Aggregated Gaussian Process Posterior}
    \label{appendix:derivation}
    Since we have
    \begin{talign*}
    \begin{pmatrix}
    y_1\\
    \vdots \\
    y_n \\
    \mathbf{w}_*^\intercal \mathbf{f}_*
    \end{pmatrix} = 
    \mathcal{N}\Bigg(
    \begin{pmatrix}
    \mathbf{w}_1^\intercal m_{\mathbf{X}_1}\\
    \vdots \\
    \mathbf{w}_n^\intercal m_{\mathbf{X}_n} \\
    \mathbf{w}_*^\intercal m_{\mathbf{X}_*}
    \end{pmatrix}
    ,  
    \begin{pmatrix}
    K +\sigma^2 I_n& K_*\\
     K_*^\intercal & K_{**}
    \end{pmatrix} 
    \Bigg),
    \end{talign*}
    where $(K)_{ij} = (\mathbf{w}_i^\intercal k_{\mathbf{X}_i\mathbf{X}_j}\mathbf{w}_j)_{ij}$, $(K_*)_i:= (\mathbf{w}_i^\intercal k_{\mathbf{X}_i\mathbf{X}_*}\mathbf{w}_*)_i$ and $K_{**}:= \mathbf{w}_*^\intercal k_{\mathbf{X}_* \mathbf{X}_*}\mathbf{w}_*$. We can thus obtain the posterior $\mathbf{w}_*^\intercal \mathbf{f}_* | y \sim \mathcal{N}(\tilde{m}_{\mathbf{X}_*}, \tilde{k}_{\mathbf{X}_*\mathbf{X}_*} )$ via standard multivariate Gaussian conditioning to get
    \begin{talign*}
        \tilde{m}_{\mathbf{X}_*} &\defeq \mathbf{w}_*^\intercal m_{\mathbf{X}_*} +  K_*^\intercal (K+\sigma^2 I_n)^{-1}(y - m_\mathbf{X}), \\ 
        \tilde{k}_{\mathbf{X}_*\mathbf{X}_*} &\defeq K_{**} - K_*^\intercal (K+\sigma^2 I_n)^{-1} K_*.
    \end{talign*}
    If elements of $\mathcal{P}(\mathcal{X})$ are explicitly known, then we would obtain 
    \begin{talign*}
    (K)_{ij} &\defeq \int_{\mathcal{X}}\int_{\mathcal{X}}  k(u,v) \text{d}\Pi_i(u) \text{d}\Pi_j(v),\\ 
    (K_*)_i&\defeq\int_{\mathcal{X}}\int_{\mathcal{X}} k(u,v) \text{d}\Pi_*(u) \text{d}\Pi_i(v) , \\ 
    K_{**}&\defeq\int_{\mathcal{X}}\int_{\mathcal{X}} k(u,v) \text{d}\Pi_*(u) \text{d}\Pi_*(v).
    \end{talign*}
    
    \section{Experiments: Additional Information}
    \label{appendix:experiments}
    In this section, we provide the details for how the data is obtained and processed. We also provide additional details for each experiment.
    
    \paragraph{Data Engineering:} We obtained the covariates via the Google Earth Engine (GEE) Python API \citep{gorelick2017google}. A Javascript-based code editor is available on the Google Earth Engine (GEE) website and allows for rapid visualisation of all the datasets within GEE. All the datasets we used are available on GEE. The key classes in GEE are \url{Image}, \url{ImageCollection} and \url{MultiPolygon}. \url{Image} stores information on a raster, or image, and \url{ImageCollection} stores a set of \url{Image} objects. \url{MultiPolygon} stores information on the geometries, or boundaries, of regions. For example, each county in our paper will have a corresponding boundary. With \url{MultiPolygon}, we can use the \url{getRegion()} method of \url{ImageCollection} to extract data, and hence pixels containing covariate data, only within the \url{MultiPolygon} geometry. We used geometries for US counties that are publicly available online, such as from \url{https://public.opendatasoft.com/explore/dataset/us-county-boundaries/table/?disjunctive.statefp&disjunctive.countyfp&disjunctive.name&disjunctive.namelsad&disjunctive.stusab&disjunctive.state_name}.
    
    GEE objects are processed on the server-side via client-side requests. However, once processed on the server side, we can directly request the data to be transferred to the client-side if the data size is moderate calling the \url{getInfo()} method from any GEE object. Another option is to download the data to Google Drive or Cloud. Further information can be found here \url{https://developers.google.com/earth-engine/tutorials/community/intro-to-python-api-guiattard}.
    
    \begin{itemize}
        \item \textbf{Masking:} We applied a 30m resolution soybean mask \citep{nass2016usda} to both the MODIS and GRIDMET datasets. Since the soybean mask is at a much lower resolution than either datasets, we "max-downsampled" the mask to each resolution by defining a value as 0 if there are no cropland pixels in the lower resolution pixel, and 1 otherwise. This can be done in GEE via the \url{Image} methods \url{reduceResolution()} and \url{updateMask()}. Once the masking is done, we can collect the data using \url{getRegion()} on each \url{MultiPolygon}, which returns a long list of datetime, longitude, latitude and covariate values, for which we can store in a \url{pandas.DataFrame} object in Python. To obtain the longitude and latitude coordinates at 500m resolution, we used a similar procedure except that the covariates are also from the soybean mask \url{ImageCollection}.
        \item \textbf{Aggregation:} We note that we downsampled MODIS to 1000m resolution for MVBAgg and 4638.3m for VBAgg, after applying the mask. As just mentioned, we also downsampled the crop mask to 500m resolution to obtain longitude and latitude coordinates. Similar to max-downsampling, we can also use mean-downsampling using  \url{reduceResolution()}. To do this, we use the mean reducer \url{ee.Reducer.mean()} as the argument inside \url{reduceResolution()}, instead of \url{ee.Reducer.max()}. This step essentially performs \textbf{Data-Agg}, as we discussed in the main section.
        \item \textbf{Data-Rep:} To match the pixels via repetition, suppose we have 2 \url{DataFrame} objects in Python, each representing a different resolution and have columns longitude, latitude and covariate values. We first take the lower resolution \url{DataFrame}, which we call \url{df1}, and use the \url{Point} object in the \url{Shapely} to create a \url{Point} for each row, using the longitude and latitude. A \url{Point} simply stores the longitude and latitude of the centre of the pixel but allows us to perform spatial operations in conjunction with the \url{geopandas} library. We then convert \url{df1} into a \url{GeoDataFrame} in \url{geopandas}. Next, given that you know the spatial resolution of \url{df2} (e.g. for 500m resolution, the spatial resolution would be 0.0089835 degrees), we create a square \url{Polygon} object for each row, representing the boundaries of each pixel, and then also convert it to a \url{GeoDataFrame} object. Lastly, we run \url{gpd.sjoin(df1, df2)}, a spatial join, that performs \textbf{Data-Rep} and returns us the required data.
        \item \textbf{Subsampling:} Due to computational constraints, during training, we have had to subsample each bag so that it is computational feasible, and this is because some bags can have up to thousands of items. Given a subsample threshold of $N$,  we randomly subsample $\min(N, N_{\text{actual}})$, where $N_{\text{actual}}$ is the actual number of elements in a bag, number of rows of $\mathbf{X}_i^l$ for a given resolution $l$.
        
    \end{itemize}
    
    All the code is provided in the supplementary material and will be open-sourced on GitHub upon publication. For each Gaussian process model, we used an additive kernel $k$, with $k^l$ being the squared-exponential kernel
    \begin{talign*}
    k(x,y)=\sigma^2\exp\left(-\frac{||x-y||_2^2}{\ell^2}\right).
    \end{talign*}
    for all $l$. One exception is that for longitude and latitude, we used the Mat\'ern-3/2 kernel:
    \begin{talign*}
    k(x,y)=\sigma^2\left( 1+ \frac{\sqrt{3}||x-y||_2}{\ell}\right)\exp\left(-\frac{\sqrt{3}||x-y||_2}{\ell}\right).
    \end{talign*}
    For all experiments we trained VBAgg and MVBAgg with 20000 iterations, $n_b=50$, $\text{lr}_{\text{Adam}}=0.001$ and $\gamma_{\text{NatGrad}}=0.1$ (the learning rate of the natural gradient). We kept the inducing points $\Z$ fixed as it did not make too much of a difference as opposed to optimising them. To enable optimising for $\Z$, it can be done by simply switching from \url{False} to \url{True} in the code. For CGP, we learned the hyperparameters via maximum likelihood using L-BFGS-B until convergence, with a maximum of 500 iterations. During training, we normalised both the features and response using \url{StandardScaler} in \url{sklearn} and unnormalised during prediction time.
    
    \paragraph{Semi-synthetic:} To reiterate, we used the "original" dataset (without subsampling) for which we have longitude-latitude (500m, downsampled), MODIS (1000m resolution, downsampled) and GRIDMET (original resolution), to generate the synthetic response.
    
    \paragraph{Crop Yield Modelling:} We obtained the crop yield data via the USDA Quickstats \citep{statsusda} user interface. We then log-transformed the yields since we choose to model log-yields. To compute the Sobol indices, denote $\alpha^l(x) = k^l_{x \Z}k_{\Z\Z}^{-1}\mathbf{m}$ and $\alpha(x) = k_{x \Z}k_{\Z\Z}^{-1}\mathbf{m}$. Then using the full dataset obtain via \url{Data-Rep}:
    \begin{talign*}
        \text{Var}_x[\tilde{m}(x)] &\approx \frac{1}{N}\sum_{j=1}^N \alpha(x_j)^2 - [\frac{1}{N}\sum_{j=1}^N \alpha(x_j)]^2 \\
        S_l &\approx \frac{1}{N(\text{Var}_x[\tilde{m}(x)]+\sigma^2)}\sum_{j=1}^N (\alpha^l(x_j))^2 - [\frac{1}{N}\sum_{j=1}^N \alpha^l(x_j)]^2 \\
        S_{lh} &\approx \frac{2}{N(\text{Var}_x[\tilde{m}(x)]+\sigma^2)}\sum_{a=1}^N (\alpha^l(x_a) - \frac{1}{N}\sum_{j=1}^N \alpha^l(x_j))(\alpha^h(x_a) - \frac{1}{N}\sum_{j=1}^N \alpha^h(x_j)).
    \end{talign*}
    \paragraph{Visualisation:} In order to visualise the pixel values, such as for Figure~\ref{fig:EVI_2counties} and Figure~\ref{fig:crop_disagg}, we can simply use the built-in \url{plot()} method of \url{GeoDataFrame}'s when each row has a \url{Polygon} to represent the pixel boundaries.
    
    \section{Model Implementation}
    \label{appendix:implementation}
    In this section, we give the key details to how the models were implemented.
    
    \paragraph{Bag Dataset Class:} For usual supervised learning datasets of the format $\{(x_i,y_i)\}_{i=1}^n$, one can simply store the covariates as a 2D array $\mathbf{X}\in\mathbb{R}^{n\times d}$. In this case, the index pointing towards the covariates and response are the same. However, for bag data, where 1 response corresponds to $N$ different samples of covariates, we would need a index for each $i$. We implement \url{BagData} that stores $\mathcal{D}:=\{(N_{i1},\ldots,N_{id},\mathbf{w}_i^1,\ldots,\mathbf{w}_i^d,\mathbf{X}_i^1,\ldots,\mathbf{X}_i^d,y_i)\}_{i=1}^n$.
    
    We index each $i$ using a \url{key}, which can be user-defined. For crop yield modelling, our key is `State-County`. To initialise \url{BagData}, we must first define a nested dictionary  \url{data_dict}:=$\{(N_{i1},\ldots,N_{id},\mathbf{w}_i^1,\ldots,\mathbf{w}_i^d,\mathbf{X}_i^1,\ldots,\mathbf{X}_i^d,y_i)\}_{i=1}^n$, where each value of the dictionary is another dictionary containing the relevant items. We can then instantiate a \url{BagData} object by calling \url{BagData(data_dict)}, which notably implements a \url{__getitem__()} method and stores the response vector $y$. Using the \url{__getitem__()} method, one can then create minibatch slices of $\mathcal{D}$ using the outputted object of \url{tf.data.Dataset.from_generator()}. One caveat is that we padded each $\mathbf{w}$ and $\mathbf{X}$ with zeros so that they are also of the same shape, defined by the maximum bag size, otherwise we cannot minibatch $\mathcal{D}$. 
    
    \paragraph{VBAgg and MVBAgg:} We implement VBAgg and MVBAgg using \url{gpflow} 2.3.0, although $>$2.2.0 should also work. We build classes \url{VBAgg} that inherits \url{gplow.models.SVGP}. We override the \url{predict_f()} and \url{elbo} methods and use a new posterior class \url{VBAggPosterior} that inherits \url{IndependentPosteriorSingleOutput}. \url{VBAggPosterior} notably has a \url{_conditional_aggregated_fused()} method that computes the aggregated Gaussian process posterior mean and variances. For \url{MVBAgg}, the implementation is very similar, except that we have multiple resolutions of kernel computations.

\end{document}